\documentclass[showpacs,amsmath,amsfonts,amssymb,preprintnumbers,superscriptaddress,twocolumn,pra]{revtex4}
\usepackage{dcolumn}
\usepackage{bm}
\usepackage[english]{babel}
\usepackage[dvips]{graphicx}
\begin{document}

\title{Quantum noise limited interferometric measurement of atomic noise: towards spin squeezing on the Cs clock
transition}
\author{Daniel Oblak}
\email{oblak@nbi.dk}
\affiliation{QUANTOP, Danish National Research Foundation Centre of Quantum Optics}
\affiliation{Department of Physics and Astronomy, University of Aarhus, DK-8000 Denmark.}
\author{Plamen G. Petrov}
\affiliation{QUANTOP, Danish National Research Foundation Centre of Quantum Optics}
\affiliation{Niels Bohr Institute, DK-2100 Copenhagen \O, Denmark.}
\author{Carlos L. \surname{Garrido Alzar}}
\affiliation{QUANTOP, Danish National Research Foundation Centre of Quantum Optics}
\affiliation{Niels Bohr Institute, DK-2100 Copenhagen \O, Denmark.}
\author{Wolfgang Tittel}
\altaffiliation[Present address, ]{University of Geneva, Group of Applied Physics, 20, Rue de l'Ecole de Medecine, CH-1211 Geneva 4, Switzerland.}
\author{Anton K. Vershovski}
\altaffiliation[Present address, ]{Ioffe Phys.-Tech. Institute, Quantum Magnetometry Lab, 26, Polytechnisheskaya, St.-Petersburg, 194021 Russia.}
\author{Jens K. Mikkelsen}
\affiliation{QUANTOP, Danish National Research Foundation Centre of Quantum Optics}
 \affiliation{Department of Physics and Astronomy, University of Aarhus, DK-8000 Denmark.}
\author{Jens L. S\o rensen}
\affiliation{QUANTOP, Danish National Research Foundation Centre of Quantum Optics}
\affiliation{Department of Physics and Astronomy, University of Aarhus, DK-8000 Denmark.}
\author{Eugene S. Polzik}
\affiliation{QUANTOP, Danish National Research Foundation Centre of Quantum Optics}
\affiliation{Niels Bohr Institute, DK-2100 Copenhagen \O, Denmark.}

\date{\today}

\pacs{42.50.Lc, 42.50.Nn, 06.30.Ft, 03.65.Ta}
\keywords{Qunatum non-demolition measurement, Spin squeezing, Atomic
  projection noise, Atomic clocks}

\begin{abstract}
We investigate theoretically and experimentally a nondestructive
interferometric measurement of the state population of an ensemble
of laser cooled and trapped atoms. This study is a step towards
generation of (pseudo-) spin squeezing of cold atoms targeted at
the improvement of the Caesium clock performance beyond the limit
set by the quantum projection noise of atoms. We calculate the phase shift and the quantum noise of
a near resonant optical probe pulse propagating through a cloud of
cold $^{133}Cs$ atoms. We analyze the figure of merit for a quantum
non-demolition (QND) measurement of the collective pseudo-spin and show that it can be
expressed simply as a product of the ensemble optical density and
the pulse integrated rate of the spontaneous emission caused by the
off-resonant probe light. Based on this, we propose a
protocol for the sequence of operations required to generate and
utilize spin squeezing for the improved atomic clock
performance via a QND measurement on the
probe light. In the experimental part we demonstrate
that the interferometric measurement of the atomic population can
reach the sensitivity of the order of $\sqrt{N_{at}}$ in a
cloud of $N_{at}$ cold atoms, which is an important benchmark towards
the experimental realisation of the theoretically analyzed protocol.
\end{abstract}

\maketitle


\section{Introduction}

The quantum features of a collective atomic magnetization were experimentally addressed for the first time in the study of quantum noise of a collective spin done by Alexandrov and Zapasskii~\cite{aleksandrov}. The relevance of this work was later on accentuated by the first observation of a quantum noise limited magnetization, i.e. linear dependence of the atomic variance on the number of atoms, by S\o rensen et. al.~\cite{sorensen}. This quantum limit, called projection noise limit, has been reached in state-of-the-art Cs atomic clocks~\cite{paris} and, by now, is the limitation towards the improvement of the clock precision. However, the projection noise would not be the limiting factor for the clock precision if one was able to increase the number of atoms used in the clock operation, but this has not been possible due to large collisional shift in cold Cs samples.

Nevertheless, it is possible to overcome the projection noise limit, as it has been demonstrated with the generation of entangled and squeezed states of two ions~\cite{squeezion}, and the creation of spin squeezed states in a cloud of cold excited Cs atoms~\cite{hald}. As a protocol for the generation of spin squeezed states, the use of a quantum nondemolition (QND) measurement has been proposed and implemented in a vapor cell in~\cite{kuzmich}, and the same kind of interaction was first proposed in~\cite{kuzmich,kuzmich1} as a mean for improving the clock performance. Another approach towards performing atomic spectroscopy below the standard quantum limit using Bose-Einstein condensates has been suggested in~\cite{kasevich2}.
 
In this paper we propose a sequence of spin rotations and QND measurements which should allow to overcome the projection noise limit for a Cs clock. The method involves generation of the coherent superposition of the two hyperfine level states, followed by a quantum nondemolition measurement of the population difference of the two states, and a sequence specific spin rotations. We present the theory of a quantum noise limited interferometer, with an atomic cloud placed in one of its arms, as a device to be used for the generation of a spin squeezed atomic sample. We report the first experiment on nondestructive interferometric measurement of the atomic population with the sensitivity approaching the quantum limit. The experimental results are obtained with an atomic ensemble in a thermal equilibrium of the two hyperfine ground states. Therefore, in the experiment we do not measure the projection noise but rather the atomic population fluctuations. The goal of the experimental part is to show that a nondestructive measurement of the atomic population with sensitivity of the order of $\sqrt{N_{at}}$, corresponding to the projection noise sensitivity, can be achieved with our methods and hence the feasibility of the method. A nondestructive measurement of the atomic level population using phase contrast imaging has been reported in~\cite{cornell}.

The paper is organized as follows. In theoretical Sec.~\ref{sec:theory}, we introduce the pseudo-spin in the Bloch sphere picture. We proceed to derive the equations governing the interaction of the probe field with the Caesium atoms, followed by an analysis of the noise contributions to the interferometric signal. The next two theoretical sections deal with the effects creating and counteracting spin squeezing and we end the theory by illustrating qualitatively how the squeezed pseudo-spin can be incorporated into the clock operation with the aim of improving its precision.

In the experimental section we start out with the description of our setup and continue to document its operational properties, with emphasis on the interferometer noise and methods to supresion unwanted noise. We present the results of a measurement of the phase shift due to the atoms in Sec.~\ref{sec:dcphase}, and in Sec.~\ref{sec:spinnoise} we show the results of a measurement of the atomic population noise of our cold atoms with the sensitivity of $\sqrt{N_{at}}$. We conclude in Sec.~\ref{sec:outlook} and present the outlook towards the future implementations and improvements of the experiment.


\section{Theoretical description}\label{sec:theory}

\subsection{Pseudo spin in the Bloch sphere representation}\label{sec:suggestion}
Let us first introduce the two level atom formalism to describe the hyperfine ground levels of Caesium. Considering a two level atom $k$ with the states $|3\rangle$ and $|4\rangle$ , the density matrix elements are $\hat{\rho}_{ij}^k=|i\rangle \langle j|_k$ ($i,j=3,4$). Note, that in atomic clocks the two levels are the $m=0$ magnetic sublevels but in much of the discussion below the magnetic state of the two levels is not important. Such a two level atom can be described in terms of a spin one half system with the pseudo spin operators defined by
\begin{eqnarray}
 \hat{\jmath}_x^{\hspace{1pt}k} &=&\frac{1}{2}(\hat{\rho}_{43}^k+\hat{\rho}_{34}^k)\ ,\nonumber \\
 \hat{\jmath}_y^{\hspace{1pt}k} &=&\frac{-i}{2}(\hat{\rho}_{43}^k-\hat{\rho}_{34}^k)\ ,\\
 \hat{\jmath}_z^{\hspace{1pt}k} &=&\frac{1}{2}(\hat{\rho}_{44}^k-\hat{\rho}_{33}^k)\ ,\nonumber
 \label{eq:pseudospin}
\end{eqnarray}
where $\hat{\jmath}_x^{\hspace{1pt}k}$, $\hat{\jmath}_y^{\hspace{1pt}k}$ and $\hat{\jmath}_z^{\hspace{1pt}k}$ are the projections of the angular momentum operator $\hat{\jmath}^{\hspace{1pt}k}$ on the $x$, $y$ and $z$ axes, respectively.

For an ensemble of $N_{at}$ atoms, we define the collective angular momentum operators by $\hat{\jmath}_x =\sum_k \hat{\jmath}_x^{\hspace{1pt}k}$ , for the $x$ component, and similarly for the other ones. These operators fulfil the angular momentum algebra $[\hat{\jmath}_i,\hat{\jmath}_j]=i\varepsilon_{ijl}\hat{\jmath}_l$, where $\varepsilon_{ijl}$ is the Levi-Civita tensor, and are useful for illustrating the evolution of the atomic quantum state using the Bloch sphere representation. This representation is obtained by plotting a vector which $x$, $y$ and $z$ components are given by the mean values of $\hat{\jmath}_x,\hat{\jmath}_y$ and $\hat{\jmath}_z$, respectively. From this picture, the conservation of $\langle\hat{\jmath}\hspace{1pt}\rangle^2=\langle\hat{\jmath}_x\rangle^2+\langle\hat{\jmath}_y\rangle^2+\langle\hat{\jmath}_z\rangle^2$, which is equivalent to the conservation of the number of atoms, is represented as a trajectory of $\langle\hat{\jmath}\hspace{1pt}\rangle$ on the surface of a sphere. The population difference between the two atomic levels is then given by the projection of $\hat{\jmath}$ on the polar axis ($\hat{\jmath}_z$), whereas the projection onto the sphere's equatorial plane gives information on the coherent superposition of the two atomic states $|3\rangle$ and $|4\rangle$.

\begin{figure*}[t!]
 \includegraphics[width=\textwidth]{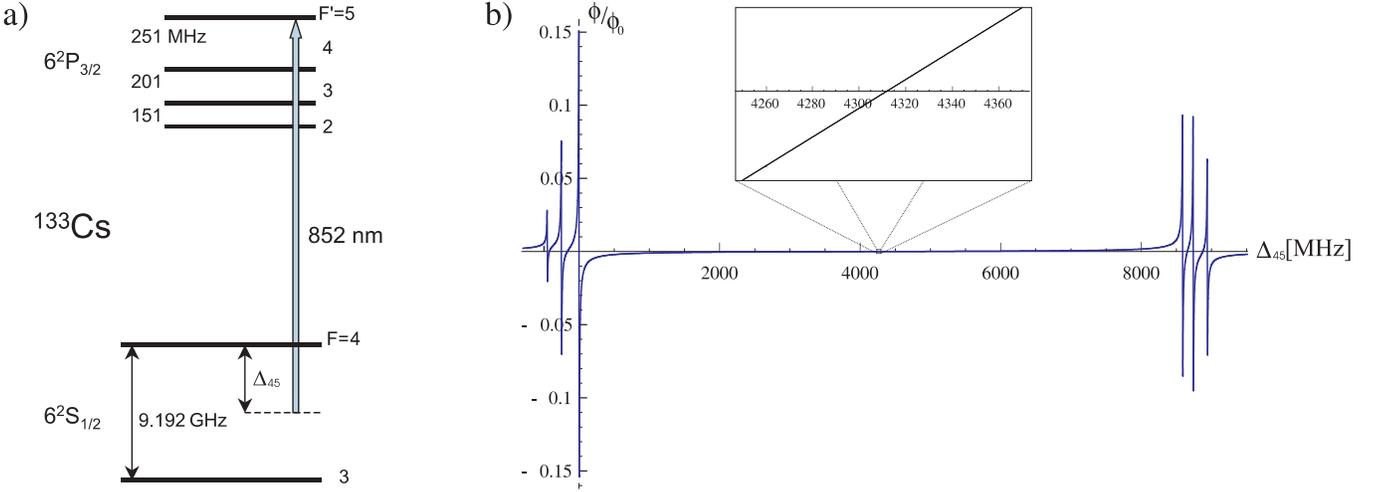}%
 \caption{(Color online) a) Diagram of the Cs hyperfine levels
 included in the $D_{2}$ line. b) Theoretically evaluated phase shift of
 the probe as a function of the detuning $\Delta_{45}$ from the
 $6S_{1/2}(F\!=4)\longrightarrow 6P_{3/2}(F'\!=5)$ transition.}
\label{fig:cslevels}
\end{figure*}


\subsection{Atomic phase shift}\label{sec:phaseshift}

\begin{figure}[t!]
 \includegraphics[width=0.45\textwidth]{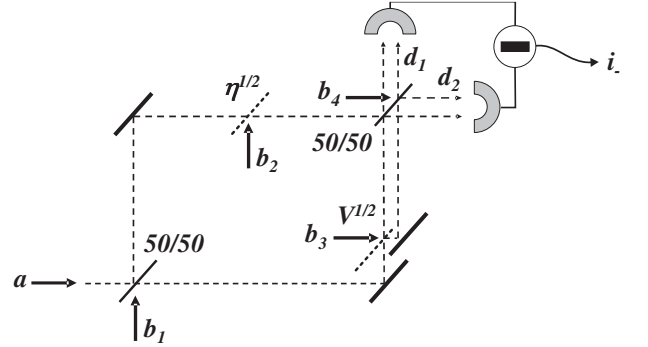}%
 \caption{(Color online) The Mach-Zehnder interferometer with loss sources and
   associated input fields indicated. The atoms are considered to be in the upper arm.}
\label{fig:inter}
\end{figure}

The initial step is to investigate how an atomic pseudo spin can influence the phase of an optical probe field near resonance on a transition between the hyperfine ground states and an excited state. To this end, we start by writing up the complex index of refraction imposed on off-resonant light by a sample of cold multilevel atoms. We consider the alkali $D$ transition $J \rightarrow J'$ between states having total electronic angular momenta $J$ and $J'$. The index of refraction is given by~\cite{sobelman}

\begin{eqnarray}
 \label{eq:indexref}
 n_\Delta-1 \!\!&=&\!\! \frac{\lambda^3}{8 \pi^2} (2J+1)\\ &&\!\!\times \sum_{F,F'} N_F (2F'+1)
 \left \lbrace \begin{matrix} J\!\!&\!\!F\!\!&\!\!I\cr F'\!\!&\!\!J'\!\!&\!\!1 \end{matrix} \right \rbrace^2 \gamma \frac{\Delta_{FF'}\!+i\gamma}{\Delta_{FF'}^2\!+\gamma^2}\ \nonumber,
\end{eqnarray}
where $I$ and $F$ are the nuclear and total atomic ground state angular momenta respectively, and the primed quantum numbers refer to the excited states. We have also introduced $N_F$ for the atomic density in the level with angular momentum $F$, $\Delta_{FF'}= \omega_{FF'}-\omega$ for the detuning of the probe light from the $F \rightarrow F'$ transition, the atomic linewidth $\gamma$ and finally the wavelength $\lambda$, assumed to be common for all transitions making up the considered $D$ line. Eq.~(\ref{eq:indexref}) is valid for a polarized probe interacting with a currently experimentally realisable unpolarized atomic ground state so that the population density in the gound state magnetic sublevel $\vert F,m_F \rangle$ is $N_{F,m_F}=N_F/(2F+1)$, and we have assumed detunings small enough to have $\vert \Delta_{FF'} \vert \ll \omega$. For the Caesium $D_2$ line ($J'=3/2$) of relevance in our experiment, we have $F=\{3,4\}$ and $F'=\{2,3,4,5\}$, as shown on Fig. \ref{fig:cslevels}a). As we will see, the phase-shift associated with the index of refraction (\ref{eq:indexref}) carries the relevant inforamtion about the $z$-component of the pseudo-spin, and can be measured using an interferometer as depicted in Fig.~\ref{fig:inter}

Eq. (\ref{eq:indexref}) is linked to the pseudo-spin, when we describe the population of the two hyperfine ground states in terms of the pseudo spin component $\hat{\jmath}_z$.

If we consider the situation where both hyperfine ground states are close to being equally populated, ($N_3=N_4$),  then in the pseudo-spin language we will have $\langle \hat{\jmath}_z \rangle=0$, and let us say only $\langle \hat{\jmath}_x \rangle=j$ with a nonzero mean value. In the Bloch sphere representation, this situation corresponds to a vector in the equatorial plane as in Fig.~\ref{fig:blochspheres}b). In this situation, the atomic variance of $\hat{\jmath}_z$ is the same as that of $\hat{\jmath}_y$ and equal to $N_{at}/4$~\cite{radcliffe,arecchi}. This can be depicted on the Bloch sphere [Fig.~\ref{fig:blochspheres}b)] by an uncertainty disk at the tip of, and perpendicular to the mean value of the angular momentum vector. It is well known \cite{kuzmich,kuzmich1} that by performing a QND measurement of $\hat{\jmath}_z$ this quantity can acquire a value more well defined than that corresponding to an ensemble of independent atoms and thus spin squeezing of the pseudo spin vector can be achieved.

In our case, the QND measurement will be performed by monitoring the optical phase shift of the  off-resonant probe interacting with our Cs atoms on the $D_2$ line. This phase shift is given by $\phi_\Delta=k_{0}l\text{Re}\{n_\Delta-1\}$, where $l$ is the physical length of our atomic sample and $k_{0}$ is the optical wavenumber. Using the Eq. (\ref{eq:indexref}), we find

\begin{eqnarray}
 \phi_\Delta\!\!&=&\!\!\frac{\phi_0}{2} \Bigg[(1\!+\beta)\! \sum_{F'=3}^5 (2F'\!+\!1) \Bigg\{ \begin{matrix} \frac{1}{2}\!&\!4\!&\!\frac{7}{2}\\[2pt] F'\!&\!\frac{3}{2}\!&\!1 \end{matrix} \Bigg \}^2 \!\frac{\gamma \Delta_{4F'}}{\Delta_{4F'}^2\!+\gamma^2} \nonumber \\
&&\!+(1\!-\beta)\! \sum_{F'=2}^4 (2F'\!+\!1) \Bigg\{ \begin{matrix} \frac{1}{2}\!&\!3\!&\!\frac{7}{2}\\[2pt]F'\!&\!\frac{3}{2}\!&\!1 \end{matrix} \Bigg\}^2 \!\frac{\gamma \Delta_{3F'}}{\Delta_{3F'}^2\!+\gamma^2} \Bigg] ,\ \ \ \
\label{eq:phaseshift}
\end{eqnarray}
where $\phi_0=\frac{\lambda^2 l N}{2 \pi}$ and we have introduced the parametrization  $N_3=N(1-\beta)/2$ and $N_4=N(1+\beta)/2$, $N$ being the overall atomic density and $\beta=(N_4-N_3)/N=\langle \hat{\jmath}_z \rangle /j$.

Using the hyperfine splittings listed in Fig.~\ref{fig:cslevels}a) and inserting the relevant values for the 6J symbols, we find by solving~(\ref{eq:phaseshift}) a zero phase shift of the probe at $\Delta_0/2\pi=4312$ MHz relative to the $F\!=4 \rightarrow F'\!=5$ transition. At this detuning the phase shifts from the two ground state transitions to the excited state hyperfine manifold cancel for equal populations ($\beta=0$), as is illustrated in Fig.~\ref{fig:cslevels}b). Therefore, at $\Delta_0$ any excursions of $\beta$ will result in an optical phase shift proportional to $\beta$ and hence information can be obtained about the collective atomic pseudo spin $\hat{\jmath}_z$ and, in particular, the quantum fluctuations of this observable~\cite{footnote2}. Since the latter are the manifestly quantum features of the collective atomic pseudo spin observable, a nondestructive measurement with the sensitivity at the level of atomic quantum fluctuations will fix $\langle\hat{\jmath}_z\rangle$ to the recorded value at the expense of measurement induced back action noise in the orthogonal $\hat{\jmath}_y$ observable. The state determination will, among other things, be limited by the accuracy of the measurement and hence it is important, for the estimation of the degree of spin squeezing achievable, to evaluate the limiting noise sources of our phase measurement. Also relevant to our study of spin squeezing is the degree to which the probe excites transitions in the atomic medium. Obviously, such excitations will partially cancel the effect of the QND measurement and therefore, it may impose limitations to the achievable degree of spin squeezing.

\subsection{Calculation of the signal-to-noise ratio for the quantum noise limited interferometry}\label{sec:theonoise}

For the monitoring of the atomic phase shift, we consider the experimental situation illustrated in Fig.~\ref{fig:inter}, where a Mach-Zehnder interferometer is placed around the atomic sample. After the interaction with the atoms the transmission of the interferometer arm is $\eta$ and the mode overlap at the second beam splitter is $\sqrt{V}$. The operator corresponding to the input probe field is designated by $\hat{a}=\alpha \hat{1}+(\hat{x}+i\hat{y})/2$, where $\alpha$ is its real mean value and, $\hat{x}$ and $\hat{y}$ are the fluctuating quadrature components. The input photon flux is then $\Phi=\alpha^2$. Similarly, we introduce the vacuum fields $\hat{b}_k=(\hat{x}_k+i\hat{y}_k)/2$ with $k=1$ to 4, mixing via the loss processes. At the final beam splitter we must consider two orthogonal spatial modes due to the non perfect mode overlap. Clearly, these modes cannot interfere optically but they will however add coherently in the detector photocurrents. For the photon fluxes impinging on the two detectors we arrive at

\begin{widetext}
\begin{eqnarray}
 \hat{d}_1^{\dagger}\hat{d}_1=\frac{1}{4}[1+\eta-2\sqrt{\eta V}\cos(\tilde{\phi})]\hat{a}^{\dagger}\hat{a}+\bigg\{\hat{a}^{\dagger}\bigg[\frac{i}{4}[1-\eta\!\!&-&\!\!2i\sqrt{\eta V}\sin(\tilde{\phi})]\hat{b}_1-\frac{1}{2}\sqrt{\frac{1-\eta}{2}}(\sqrt{V}e^{-i\tilde{\phi}}-\sqrt{\eta})\hat{b}_2 \nonumber \\
&&\quad \qquad -\frac{i}{2}\sqrt{\frac{\eta(1-V)}{2}}\hat{b}_3+\frac{1}{2}\sqrt{\frac{1- V}{2}}e^{-i\tilde{\phi}}\hat{b}_4\bigg]+h.c.\bigg\}\ ,
 \label{eq:i1} \\[5pt]
\hat{d}_2^{\dagger}\hat{d}_2=\frac{1}{4}[1+\eta+2\sqrt{\eta V}\cos(\tilde{\phi})]\hat{a}^{\dagger}\hat{a}+\bigg\{\hat{a}^{\dagger}\bigg[\frac{i}{4}[1- \eta\!\!&+&\!\!2i\sqrt{\eta V}\sin(\tilde{\phi})]\hat{b}_1+\frac{1}{2}\sqrt{\frac{1-\eta}{2}}(\sqrt{V}e^{-i\tilde{\phi}}+\sqrt{\eta})\hat{b}_2\nonumber \\
 &&\quad \qquad +\frac{i}{2}\sqrt{\frac{\eta(1-V)}{2}}\hat{b}_3-\frac{1}{2}\sqrt{\frac{1- V}{2}}e^{-i\tilde{\phi}}\hat{b}_4 \bigg]+h.c.\bigg\}\ ,
\label{eq:i2}
\end{eqnarray}
\end{widetext}
where the phase difference $\tilde{\phi}=2 \pi \big(\int_{ref} n(L) dL - \int_{probe} n(L) dL \big)/\lambda$ is $2\pi /\lambda$ times the difference of the interferometer arms' optical path length, i.e., the integral of the index of refraction over the respective arm. From this expression of $\tilde{\phi}$ it is clear that the phase can shift because of a change in either the path lengths  or the index of refraction in one of the arms, or because of a shift of the wavelength of the probe. Moreover, in the probe arm the atoms can change the index of refraction and induce a phaseshift $\phi_{\Delta}$, so that we write $\tilde{\phi}=\phi+\phi_{\Delta}$ for the total phase shift. The last contribution provides the monitoring of the atoms, while the first three add noise to the measurement.

Now, we find the mean photocurrent difference to be
\begin{equation}
 \langle \hat{\imath}_- \rangle=\left\langle \hat{d}_1^{\dagger}\hat{d}_1 \right\rangle -   \left\langle \hat{d}_2^{\dagger}\hat{d}_2 \right\rangle=\alpha^2 \sqrt{\eta V}   \cos(\tilde{\phi})\ ,
\label{eq:imin}
\end{equation}
in units of elementary charge. The visibility of our interference fringe is found from the single detector photocurrent and is given by

\begin{equation}
 \mathcal{V}=\frac{\left\langle \hat{d}_1^{\dagger}\hat{d}_1 \right\rangle_{\tilde{\phi}=\pi}- \left\langle \hat{d}_1^{\dagger}\hat{d}_1 \right\rangle_{\tilde{\phi}=0}}{\left\langle  \hat{d}_1^{\dagger}\hat{d}_1 \right\rangle_{\tilde{\phi}=\pi}+\left\langle  \hat{d}_1^{\dagger}\hat{d}_1 \right\rangle_{\tilde{\phi}=0}}=\frac{2\sqrt{\eta V}}{1+\eta}\ .
\label{eq:visibility}
\end{equation}
In the symmetric case, where $\eta=1$, this reduces to $\mathcal{V}=\sqrt{V}$ as expected.

The fluctuating part of $\hat{\imath}_-$ is now calculated by linearizing around the mean value, $\delta \hat{\imath}_- =\hat{\imath}_- - \left\langle \hat{\imath}_- \right \rangle$, and remembering that only $\hat{a}^{\dagger}\hat{a}$ has nonzero mean. As a result we find
\begin{widetext}
\begin{eqnarray}
 \frac{\delta \hat{\imath}_-}{\alpha}&=&-\sqrt{\eta V}[\cos(\tilde{\phi}) \hat{x}+\sin(\tilde{\phi}) \hat{x}_1]-\sqrt{\frac{V(1-\eta)}{2}}[\cos(\tilde{\phi})\hat{x}_2+\sin(\tilde{\phi}) \hat{y}_2]\nonumber \\
&&-\sqrt{\frac{\eta(1- V)}{2}}\hat{y}_3-\sqrt{\frac{1-V}{2}}[\cos(\tilde{\phi}) \hat{x}_4+\sin(\tilde{\phi}) \hat{y}_4]\ .
\label{eq:flucimin}
\end{eqnarray}
\end{widetext}

All the field operators in (\ref{eq:flucimin}) are uncorrelated and consequently for a coherent state input all operators contribute with $2B$, where $B$ is the bandwidth of our measurement~\cite{yarivqe}. From this, we find $(\delta i_-)^2_{coh}=B\alpha^2(1+\eta)$, which is just $2B$ times the total photon flux, $\Phi$, transmitted through the interferometer.

Let's assume that there are no atoms in the probe arm so that $\phi_{\Delta}=0$. To be sensitive to small phase shifts, we use a second laser far away from the atomic resonance to lock the interferometer at the side of the interference fringe. With this procedure applied to the system, we set the residual phase $\phi$ equal to $\pi(1/2+m)$ with $m=0,\pm 1,...$, which has the following consequences: Even if our input state is not coherent, or in other words, we are probing the atoms using a noisy laser, we will find the \emph{amplitude} noise of the probe laser [the first term in (\ref{eq:flucimin})] to be considerably suppressed due to the balanced detection. However, the laser \emph{phase} noise will remain important. We model this noise as an excess noise of the vacuum inputs $\hat{x}_1$ and $\hat{y}_2$, interfering with the probe and contributing with a variance $(1+\mathcal{N})$ relative to the vacuum state noise while $\hat{y}_3$, $\hat{x}_4$ and $\hat{y}_4$ still are at the vacuum noise level.

Considering now the presence of  atoms, their contribution to the phase noise is denoted $(\delta\phi_{\Delta})^2$, which like the laser phase noise is normalized to the probe vacuum noise level. Incorporating these values into (\ref{eq:flucimin}) and taking into account the transmission of the interferometer, we arrive at
\begin{equation}
 \frac{(\delta i_-)^2}{\Phi}=2B+V\Phi\left [\mathcal{N}+\left (\frac{\eta}{1+\eta} \right )^2  (\delta\phi_{\Delta})^2 \right ]\cos^2(\phi_{\Delta})\ ,
\label{eq:internoise}
\end{equation}
again in units of quantum noise of the transmitted probe and for the case $\tilde{\phi}=\pi(1/2+m)+\phi_{\Delta}$.

The excess noise $\mathcal{N}$ can be suppressed by operating the interferometer in the white light position since
\begin{equation}
 \mathcal{N}=\frac{\delta\omega^2}{(\delta\omega^2)_q} (k_{0}\Delta L)^2,
\label{eq:nufac}
\end{equation}
where $\Delta L$ is the optical path difference between the two arms of the interferometer, $\delta\omega^2$ is the laser frequency noise and $(\delta\omega^2)_q=\omega_0^2/\alpha^2$ is the quantum level of frequency noise.

We can also straightforward generalize Eq.~(\ref{eq:internoise}) to detectors with less than unity quantum efficiency, $\epsilon$. In this case we get
\begin{equation}
 \frac{(\delta i_-)^2}{\epsilon\Phi}=2B+\epsilon V\Phi\left [\mathcal{N}+\left  (\frac{\eta}{1+\eta} \right )^2 (\delta\phi_{\Delta})^2 \right ]\cos^2(\phi_{\Delta})\ .
\label{eq:internoiseqe}
\end{equation}
Having addressed the optical noise contributions we will now turn to consider the atomic imprint on the probe phase noise.


\subsection{Spin squeezing}\label{sec:spinsqueezing}
In the context of spin squeezing a high ratio of atomic spin noise to optical quantum noise is desired. Intuitively this is clear because a higher signal to noise yields more knowledge of the atomic spin observable, hence it becomes better defined and higher degree of squeezing of that observable is achieved.

Assuming that we have a white light interferometer, the noise contributions of relevance here are the quantum fluctuations $(\delta i)_p ^2$ of the phase of the probe pulse and the electronic noise $(\delta i)_e ^2$ of our photodetectors. The probe pulse is characterized by the duration $\tau$ and the photon number $\Phi \tau$. Since a highly coherent laser is being used to generate the pulse, we assume it to be Fourier limited~\cite{yarivqe}, $2\pi B\tau=1$, and then
\begin{equation}
 (\delta i)_p ^2=2B\frac{\epsilon\Phi\eta}{2}= \frac{\epsilon\Phi\eta}{2\pi\tau}\ ,
\label{eq:probenoise}
\end{equation}
where we have used the photon flux detected from the probe arm $\eta\Phi/2$ as reference for the shot noise level.

The electronic noise can be described by the noise equivalent power (NEP), $P_e$ , so that
\begin{equation}
 (\delta i)_e ^2=(P_e \lambda/hc)^2 /2\pi\tau\ ,
\label{eq:elecnoise}
\end{equation}
$h$ being Planck's constant.

If we ignore the electronic noise [$(\delta i)_e ^2 \ll (\delta i)_p ^2$] that is only relevant when feedback schemes are involved, we find that the signal to noise ratio of the measurement is given by
\begin{equation}
 \kappa^2=\epsilon V \frac{2\eta}{(1+\eta)^2}\pi\Phi\tau (\delta\phi_{\Delta})^2 \cos^2(\phi_{\Delta})\ ,
\label{eq:kappadef}
\end{equation}
which is related to the degree of spin squeezing ~\cite{duan}, as described below.

The atomic contribution to the phase noise is computed from Eq.~(\ref{eq:phaseshift}). We assume that the atoms initially are prepared in a coherent spin state for which $(\delta \beta)^2_{coh}=\langle \delta \hat{\jmath}_z^2 \rangle_{coh} / j^2=N_{at}^{-1}$, where $N_{at}$ is the number of atoms within the probe volume. Hence, we find that
\begin{equation}
 (\delta\phi_{\Delta})^2=\left (\frac{\lambda^2 \mathcal{D}(\Delta)}{4\pi A} \right )^2 N_{at}\ ,
\label{eq:atomnoise}
\end{equation}
where $A$ is the probe beam cross sectional area found from the probe beam waist $w_0$ as $\pi w_0^2/2$ and we have defined the detuning function
\begin{eqnarray}
 \mathcal{D}(\Delta)\!&=&\!\sum_{F'=3}^5 (2F'\!+1) \Bigg\{ \begin{matrix} \frac{1}{2}\!&\!4\!&\!\frac{7}{2}\\[2pt] F'\!&\!\frac{3}{2}\!&\!1 \end{matrix} \Bigg\}^2 \frac{\gamma \Delta_{4F'}}{\Delta_{4F'}^2+\gamma^2} \nonumber \\
&&\!-\sum_{F'=2}^4 (2F'\!+1) \Bigg\{ \begin{matrix}\frac{1}{2}\!&\!3\!&\!\frac{7}{2}\\[2pt]F'&\!\frac{3}{2}\!&\!1 \end{matrix} \Bigg\}^2 \frac{\gamma\Delta_{3F'}}{\Delta_{3F'}^2+\gamma^2}\ .\ \ \ \
\label{eq:detunfct}
\end{eqnarray}

Finally, we find the ratio of the phase noise from the atoms to the quantum phase noise to be
\begin{equation}
 \kappa^2=\left (\frac{\lambda^2 \mathcal{D}(\Delta)}{4A} \right )^2
 \frac{2\eta}{(1+\eta)^2}\frac{\epsilon V N_{at} \Phi\tau}{\pi} \cos^2(\phi_{\Delta})\ .
 \label{eq:signoise}
\end{equation}

For our pulsed measurement we integrate $\hat{\imath}_-$ over the pulse duration and analyze the statistics of collections of pulses. This sets an upper limit to the frequency of the fluctuations, that can be observed, at approximately $\tau^{-1}$. The lower limit is simply set by the time over which we collect the integrated pulses. Since the atomic noise spectrum is not white, we stress that (\ref{eq:signoise}) is only valid in as much as we match our pulse spectrum to cover the atomic noise spectrum.

With respect to the spin squeezing, we will be using the following definition taken from~\cite{wineland}
\begin{equation}
 \xi = (\delta \beta)^2 N_{at} = \langle \delta \hat{\jmath}_z ^2 \rangle  N_{at} / j^2\ ,
\label{eq:defspinsqueez}
\end{equation}
where $\xi = 1$ for a coherent state, $\xi < 1$ for a squeezed state and $\xi = \infty$ for a thermal state. This is not the only way to define the spin squeezing parameter, but this definition characterizes the quality of the state in a spectroscopic measurement, i.e., it is a measure for the increased sensitivity to rotation in the squeezed direction on the Bloch sphere~\cite{wineland}.

The degree of spin squeezing can be shown \cite{duan,internal} to be related to $\kappa$ through
\begin{equation}
 \xi = \frac{1}{1+\kappa^2} \ ,
\label{eq:deftheta}
\end{equation}
and thus, the squeezing imprinted by the measurement onto the z-component of the spin is
\begin{equation}
 (\delta j_z)^2_{sq} = \xi (\delta j_z)^2_{coh} =\frac{1}{4} \frac{N_{at}}{1+\kappa^2}\ ,
\end{equation}
so that, in order to perform a good QND measurement and hence, to achieve a high degree of spin squeezing, we must have $\kappa$ large compared to unity. It is natural therefore to call $\kappa$ the figure of merit of the QND interaction. Below we show how $\kappa$ can be expressed via easily accessible experimental parameters.
It is important to note that as long as the spontaneous emission rate over a pulse is negligible the minimal uncertainty state will be preserved by the phase shift measurement~\cite{kuzmich1,briannature,bookchapter}, and in this case the variance of the conjugate spin component will become $(\delta j_y)^2_{sq} =\frac{1}{4} N_{at}(1+\kappa^2)$.

So far, we have ignored the electronic noise in the above calculations. It is however important if feedback schemes should be applied in order to either enhance the spin squeezing~\cite{wiseman}, or if we wish, to rotate the mean spin direction according to our measurement with the goal of obtaining a specific spin squeezed state. The latter is relevant if the spin state should be employed in, e.g., atomic clocks, where the a Bloch vector in the equatorial plane is desired~\cite{clock}.

The degree of spin squeezing can be measured experimentally by sending pairs of probe pulses through the atomic sample, integrating these pulses and storing the resulting areas, $a_1$ and $a_2$. The variances $\delta a_1^2=\delta a_2^2$ will set the level of the atomic quantum noise, while the variance of the pulse difference $\delta (a_1-a_2)^2$ will yield information of interatomic correlations created by the quantum measurement of the first pulse~\cite{briannature}. If we have created a spin squeezed ensemble it will reveal itself via the reduced variance
\begin{equation}
 \delta (a_1-a_2)^2<\delta a_1^2+\delta a_2^2=2 \delta a_1^2=2 \delta a_2^2\ .
 \label{eq:spsqcrit}
\end{equation}

The quantum nature of the atomic fluctuations can be verified by showing that $\delta a_1^2$ and $\delta a_2^2$ are equal and grow linearly with $N_{at}$. From this experimental considaration it is already clear that the sensitivity of the detection apparatus must be large enough for the $N_{at}$ noise to be detected.


\subsection{The relation between the figure-of-merit and atomic decoherence}\label{sec:probeexc}

The effect counteracting the spin squeezing is the incoherent transfer of atoms from the spin squeezed state via optical excitation and spontaneous emission to a mixed ground state. This excitation happens with a pulse integrated rate $p_e$
\begin{equation}
 p_e =\frac{\sigma(\Delta)\Phi \tau}{A}\ ,
 \label{eq:defpe}
\end{equation}
where the absorption cross section for the probe is $\sigma(\Delta)=(\lambda^2/3\pi)\mathcal{L}(\Delta)$ with the linewidth function given as
\begin{eqnarray}
\mathcal{L}(\Delta)\!&=&\!\sum_{F'=3}^5 (2F'\!+1) \Bigg\{ \begin{matrix} \frac{1}{2}\!&\!4\!&\!\frac{7}{2}\\[2pt] F'\!&\!\frac{3}{2}\!&\!1 \end{matrix} \Bigg\}^2\frac{\gamma^2}{\Delta_{4F'}^2+ \gamma^2} \nonumber \\
&&\!+\sum_{F'=2}^4 (2F'\!+1) \Bigg\{ \begin{matrix} \frac{1}{2}\!&\!3\!&\!\frac{7}{2}\\[2pt]F'&\!\frac{3}{2}\!&\!1 \end{matrix} \Bigg\}^2\frac{\gamma^2}{\Delta_{3F'}^2+ \gamma^2}\ . \ \ \ \
\label{eq:abscross}
\end{eqnarray}

The above cross section assumes both ground states having equal populations and we find that the pulse integrated excitation rate is related to $\kappa$ through
\begin{equation}
 \kappa^2=\epsilon V \frac{\eta}{(1+\eta)^2}\frac{\lambda^2}{4A}\frac{\mathcal{D}(\Delta)^2}{\mathcal{L}(\Delta)}\cos^2(\phi_{\Delta})N_{at }p_e\ .
\label{eq:kappaalphape}
\end{equation}

To highlight the relevant physical parameters this equation can be cast into a compact form. If we assume the visibility $V$, quantum efficiency $\epsilon$ as well as the transmission of the interferometer $\eta$ all equal to 1, then in the limit of large detunings $\Delta \gg \gamma$ we have $\mathcal{D}(\Delta)^2/\mathcal{L}(\Delta)\approx 1$ and the DC-phase shift becomes negligible so that also $\cos(\phi_{\Delta})\approx 1$, and Eq.~(\ref{eq:kappaalphape}) simplifies to
\begin{equation}
 \kappa^2=\frac{\pi}{8}\alpha_0 p_e\ ,
\label{eq:simplekappa}
\end{equation}
where we have introduced $\alpha_0$, the atomic optical density on resonance. It is clear that, in order to achieve strong spin squeezing we need a large $\alpha_0$ since we wish to keep $p_e$ small to maintain the nondemolishing character of the measurement.

Following the excitation the atoms decay spontaneously to a spherical spin state, i.e., a state characterized by zero expectation value of all spin components $\langle \hat{\jmath}_i\rangle =0$ ($i = x,y,z$). This dissipative evolution of the atomic state leads to the additional noise into the final state, resulting in less efficient spin squeezing. For small $p_e$ the degradation is of the order of $p_e$. A detailed analysis of the spin state resulting from the QND measurement and a certain amount of excitation is presented in~\cite{hammerer}.

\subsection{Spin squeezing in clock operation}

\begin{figure}[t!]
\includegraphics[width=0.39\textwidth]{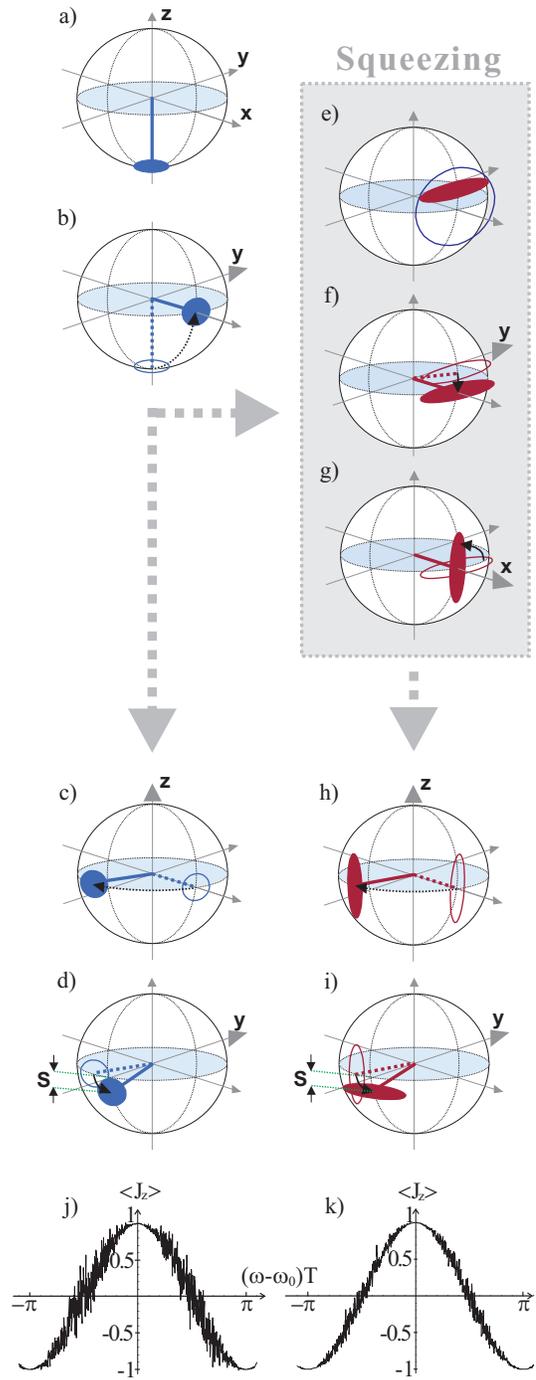}
\caption{(Color online) The Bloch-Sphere scheme for the state preparation (a,b)
  QND measurement (e,f,g) and the Ramsey spectroscopy for a coherent
  state (c,d) and a spin squeezed state (h,i), with the corresponding
  simulations of the projection noise on the Ramsey fringe, (j) and (k) respectively.}
\label{fig:blochspheres}
\end{figure}

In the preceeding sections we have shown how a non-destructive optical phase shift measurement reduces the noise of the atomic pseudo-spin $z$-component. We will now put this QND measurement into the context of the clock operation, which is discussed qualitatively in the Bloch sphere representation. More detailed accounts of the atomic clock operation in this picture can be found in \cite{wineland}.

The full protocol including the standard clock sequence can be viewed as follows: Initially we use optical pumping to prepare the atoms in a coherent spin state with $\langle \hat{\jmath}_z \rangle = -N_{at}/2$ and $\langle \hat{\jmath}_x \rangle = \langle \hat{\jmath}_y \rangle = 0$, as shown in Fig.~\ref{fig:blochspheres}a). This is the situation where the atomic variances of $\hat{\jmath}_x$ and $\hat{\jmath}_y$ are both equal to $N_{at}/4$ as depicted by the uncertainty disc on the Bloch sphere [Fig.~\ref{fig:blochspheres}a)]. Next step is to apply a classical $\pi/2$ pulse using an RF-magnetic field, which corresponds to the first $\pi/2$ pulse in the Ramsey spectroscopy sequence. This pulse brings the angular momentum vector to the equatorial plane as illustrated on Fig.~\ref{fig:blochspheres}b). For a standard atomic clock the spin would be allowed to precess in the equatorial plane of the Bloch sphere [Fig.~\ref{fig:blochspheres}c)] until the second $\pi/2$ pulse in the Ramsey sequence is applied [Fig.~\ref{fig:blochspheres}d)] and the atomic population difference is detected. Instead of this, we proceed from the state depicted on Fig.~\ref{fig:blochspheres}b) to perform a QND measurement of the population difference $\hat{\jmath}_z$ using an optical field. As argued above, this measurement reduces the uncertainty of the operator $\hat{\jmath}_z$ at the expense of an increased uncertainty of $\hat{\jmath}_y$ while preserving the minimal uncertainty state and so, the atomic sample is prepared in a spin squeezed state. This is the state illustrated in Fig.~\ref{fig:blochspheres}e). 

The degradation of the spin squeezing that can be caused by spontaneous emission will cause an increase in the size of the uncertainty disc, as well as  reduce the length of the Bloch vector leading to the loss of contrast in the clock signal. However, both effects are of the order of the pulse integrated rate of spontaneous emission which can be kept small, simply by letting the probe be far detuned.

Additionally, we must update the atomic state based on the result of the QND measurement~\cite{wiseman}. Depending on the outcome of the measurement the mean value of the pseudo spin vector will be shifted away from $\langle \hat{\jmath}_z\rangle=0$. This deviation will be corrected for by application of a short RF-pulse that will shift the vector back into the equatorial plane of the Bloch sphere [Fig.~\ref{fig:blochspheres}f)]. 

Since for the Caesium clock it is important to have reduced noise during the precession of the phase component $\hat{\jmath}_y$ in the equatorial plane, we rotate the pseudo spin vector around the $x$ axis with a $\pi/2$ pulse [Fig.~\ref{fig:blochspheres}g)], which effectively interchanges the $\hat{\jmath}_y$ and $\hat{\jmath}_z$ components. For the $\pi/2$ rotation to revolve around the $x$-axis, the RF-magnetic field, applied in this step, must be phase-shifted by $\pi/2$ w.r.t. that the of the first Ramsey pulse [Fig. \ref{fig:blochspheres}b)]. The following steps, represented by Figs.~\ref{fig:blochspheres}h)~and~\ref{fig:blochspheres}i), correspond to the standard method for Ramsey spectroscopy in the clock operation, but now using a spin squeezed state.

If we compare Figs.~\ref{fig:blochspheres}j)~and~\ref{fig:blochspheres}k) we clearly see that we would gain in signal to noise ratio of the spectroscopy signal from the projection noise limited measurement of the population difference ($\hat{\jmath}_z$) in the clock transition~\cite{wineland}, performed at $(\omega-\omega_0)T=\pm\pi/2$.


\section{Atomic noise measurements}

In this part we aim at showing that our apparatus has the sufficient sensitivity to measure the atomic noise and along the way we analyze the various considerations and precautions necessary for reaching the goal.

\subsection{Magneto-optical trap}

The atomic sample is prepared in a standard six beam Cs magneto-optical trap (MOT). We are able to trap around $3\times 10^{8}$ atoms, when loading the MOT from background Cs vapor with sufficiently high partial pressure (around $10^{-7}$ mbar). The red detuning of the trapping laser is set to 15 MHz. The cloud volume is approximately $6\times 10^{-3}$ cm$^3$, which at best can yield a resonant optical density of 13. Due to high background pressure the lifetime of the trap is only around 20 ms. This short lifetime is convenient for acquiring statistical data because it allows for quick refreshing of the atomic sample.


\subsection{Frequency locking of the probe laser}

\begin{figure}[t]
\includegraphics[width=0.45\textwidth]{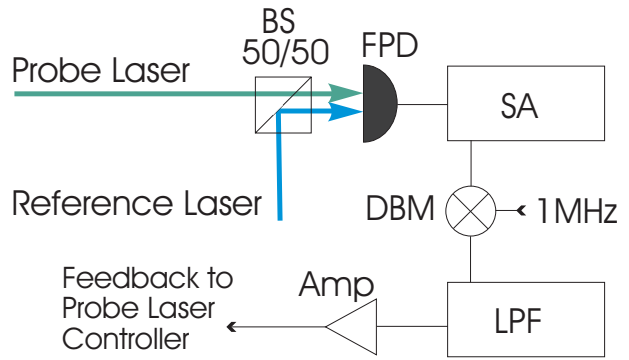}
 \caption{(Color online) Experimental setup employed to lock the probe beam. The
elements included in the sketch are: BS - 50/50 beam splitter; FPD
- fast photodetector; SA - spectrum analyser; DBM - double
balanced mixer; LPF - low pass filter; Amp - amplifier.}
\label{fig:setex}
\end{figure}

The probe laser is locked and blue detuned from the atomic transition $6S_{1/2}(F\!=4) \rightarrow 6P_{3/2}(F'\!=5)$ by a specific detuning $\Delta$ variable from a few MHz to a few GHz. The experimental setup used to lock the probe laser in this way is shown in Fig.~\ref{fig:setex}.

Two lasers with a specific relative frequency separation can be locked in a number of ways ~\cite{laserlock,laserlock1}. We use a reference laser locked to the atomic transition $6S_{1/2}(F\!=4) \rightarrow 6P_{3/2}(F'\!=5)$ by FM saturation spectroscopy. The reference and a fraction of the probe laser beams are mixed at the beam splitter $BS$ and their beat note is measured using a Newport fast photodetector (model 1480) with 15 GHz bandwidth. The produced RF signal has a frequency component corresponding to the probe detuning from the above specified atomic transition. This signal is monitored with the spectrum analyzer $SA$, as shown in the figure. From the RF signal obtained in such a way, we can generate an error signal to lock the probe laser with the desired detuning from the atomic transition. To accomplish that, the current of the probe laser is FM modulated with a modulation depth of 1\% using a 1 MHz sinusoidal waveform that is also utilized as a local oscillator for the double balanced mixer $DBM$ in Fig.~\ref{fig:setex}. The output from the mixer is a DC signal that is low pass filtered and amplified before feeding it back to the probe laser controller.


\begin{figure*}[t]
\begin{center}
\includegraphics[width=\textwidth]{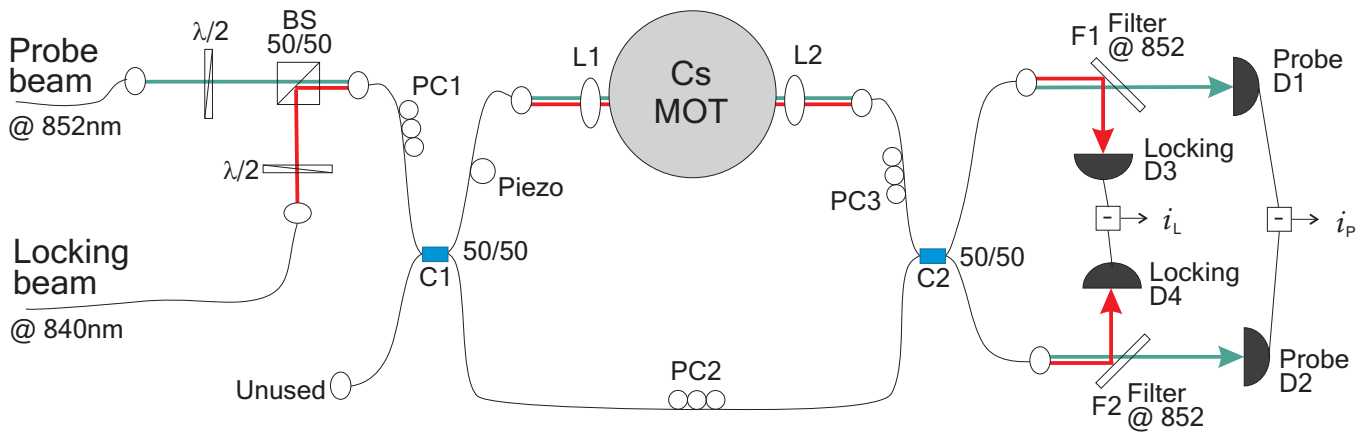}
\end{center}
\caption{(Color online) Sketch of the setup of the interferometer with following
elements: BS - 50/50 beam-splitter; C1 \& C2 - 50/50 fiber
couplers; PC1, PC2 \& PC3 - fiber polarization
controllers; L1 \& L2 - achromatic lenses; F1 \& F2 - interference
filters transmitting @ 852 nm; D1 \& D2 - Hamamatsu low noise, high
gain photodiodes; D3 \& D4 - photodetectors; and several half
wave plates $\lambda /2$ and collimating lenses for fiber
coupling. $i_L$ is the locking signal, whereas $i_- = i_1 - i_2$ is the probe signal.}
\label{fig:interf}
\end{figure*}

\subsection{White light fiber interferometer}

The interferometer as shown on Fig.~\ref{fig:interf} is a Mach-Zehnder type made of single mode optical fibers. The motivation for using fibers instead of free space propagating beams is the enhanced mechanical stability as well as excellent mode overlap of the interfering beams in single mode fibers. The field in the input fiber enters a 50/50 coupler $C1$ and is split into a reference arm and a probe arm surrounding the atoms to be probed. The field in the probe arm exits the fiber and with the lens $L1$ it is focused at the center of the MOT with a beam waist of $20~\mathrm{\mu m}$. After passing through the atoms, it again enters the fiber and is combined with the field from the reference arm at the second 50/50 coupler $C2$.

Since we use a non polarization maintaining fiber, the field polarization can evolve differently in the two arms. Thus, in order to achieve maximal interference visibility, we include polarization controllers $PC$ to match the field polarization from the two arms at the second coupler $C2$. We note that the coupling efficiency $\eta$ through the air-gap containing the MOT is 30 \%. With the fiber couplers providing a nearly perfect mode overlap of the probe and the reference field, i.e., $\sqrt{V} \approx 1$, the visibility becomes $\mathcal{V}=85\%$.

The pulsed probe signal is detected with a balanced detection scheme~\cite{konstanz} using the low noise photodiodes $D1$ and $D2$. From the integral of the photocurrent $i_-$ over the pulse duration, we extract the area which corresponds to the difference of the signals from the two arms. The mean value of the difference gives the DC-phase shift $\tilde{\phi}$, and the variance gives information about the phase fluctuations.


\subsubsection{Locking the Interferometer}

To reduce thermal and acoustic drifts of the interferometer, we lock it by means of an  off-resonant CW laser that propagates through the interferometer simultaneously with, and in the same direction as the probe beam. The locking beam is several nm away from the atomic resonance and therefore it is not affected by the cold atoms.  At the output the locking beam and the probe beam are separated by the interference filters $F1$ and $F2$. The balanced detectors $D3$ and $D4$ provide an error signal which controls the piezo adjusting the length of the probe arm.

In order to cancel the amplitude noise term in the Eq.~(\ref{eq:flucimin}), the interferometer needs to be locked so that when the cold atoms are absent ($\phi_{\Delta}=0$) the interference signal for the probe is at half fringe, i.e., $\phi=\pi(1/2+m)$ with $m=0,\pm 1,...$. However in this position the interference signal is also most sensitive to the phase noise of lasers. We use semiconductor lasers that are characterized by strong phase fluctuations with a wide band of frequencies for both probing and locking~\cite{dandridge}. We have measured their linewidth to be approximately 500kHz. As discussed in Sec.~\ref{sec:theonoise}, in order to suppress the effect of the phase noise we lock the interferometer at the white light position, corresponding to a nearly zero path length difference~\cite{dandridge}. We use a regular broadband LED as a white light source to determine \emph{roughly} the white light position.


\subsubsection{Shot noise limited interferometer}\label{sec:lightnoise}

\begin{figure}[t]
\begin{center}
\includegraphics[width=0.4\textwidth]{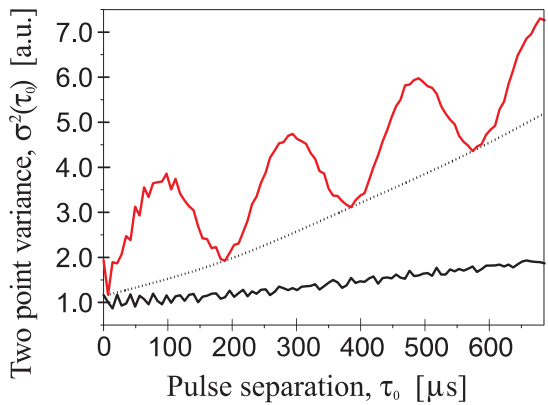}
\end{center}
\caption{(Color online) The two point variance extracted from
 measurements done with pulse separation $\tau_0$ = 20 $\mathrm{\mu
 s}$. Lower trace: Interferometer in white light position with probe laser locked and the
 variance increases on larger timescales due to the 50 Hz line
 noise. Upper trace: Interferometer out of white light position
 with probe laser frequency modulated at 5 kHz, which is directly reflected in the variance
 by a 200 $\mu$s oscillation period. Additional phase noise from the
 laser raises the level of the minima
 ($\cdot$$\cdot$$\cdot$$\cdot$$\cdot$$\cdot$) with respect to the
 lower trace, because the laser is not phase locked and the
 interferometer is not in the white light position.}
\label{fig:2pvariance}
\end{figure}

With all the measures for removing undesirable noise contributions implemented, we now try to gauge their effectivenes.
To get a measure of the instability or noise of the interferometer signal we measure the variation of the area from one pulse ($a_{i}$) to another ($a_{i+1}$). This is done by determining the two point variance~\cite{footnote1} $\sigma^2(\tau_0)$ which can be defined to be

\begin{equation}\label{eq:2pvar}
 \sigma^2(\tau_0) = \frac{1}{2(M-1)} \sum_{i=0}^{M-1}
 (a_{i+1}-a_{i})^2\ ,
\end{equation}

where $\tau_0$ is the temporal pulse separation and $M$ is the number of pulses in our measurement. The pulse sequences used to compute $\sigma^2(\tau_0)$ are composed of several thousand pulses each of 2 $\mathrm{\mu s}$ duration. From these sequences we can extract the two point variance on timescales comparable with the pulse duration and up to two orders of magnitude in $\tau$.  The results corresponding to this measurement are shown in Fig.~\ref{fig:2pvariance}.

If the interferometer noise would be purely white, the two point variance would stay constant on all timescales. Naturally, temperature drifts would cause the variance to rise on larger timescales than those we measure. We observe that the two point variance fluctuates with a period of 200 ms corresponding to 50 Hz line noise. On the scale of Fig.~\ref{fig:2pvariance}, this is seen as a slow rise of the curves. However, on the $\mathrm{\mu s}$ timescale this line noise is of no importance. One interesting feature is seen from the upper trace of Fig.~\ref{fig:2pvariance}. By modulating the laser diode current and thereby, the frequency of the probe laser, we see that $\sigma^2(\tau_0)$ oscillates with a period corresponding to the modulation frequency. This in turn means that the interferometer is sensitive to frequency changes and therefore, isn't exactly in the white light position. By adjusting the path length difference of the interferometer arms it is possible to come in to the white light position and consequently, the oscillation disappears from the two point variance as seen on the lower trace of Fig.~\ref{fig:2pvariance}. This method ensures a \emph{fine} white light alignment to within 10 $\mu$m, so that we can have $m \lesssim 2$ and fulfill the condition $\phi \lesssim 5\pi/2$. Since it is of less concern what is the exact value of $m$, for the sake of simplicity we assume it to be zero.

\begin{figure}[t]
\begin{center}
\includegraphics[width=0.4\textwidth]{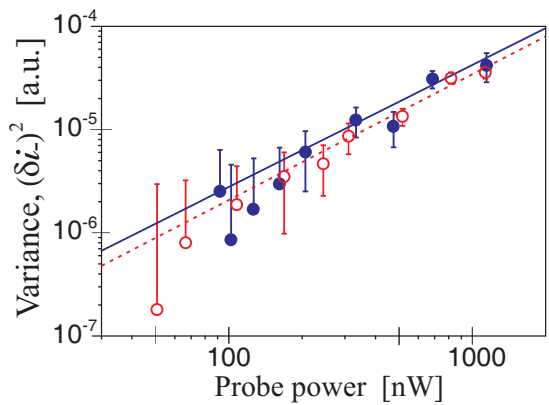}
\end{center}
\caption{(Color online) Noise in the amplitude $x$ ($\circ$) and phase $y$
  ($\bullet$) quadratures of the probe light. Fits to the
  data in log-log scale give slopes of $1.2\pm 0.2$ (-~-~-) for the amplitude and
  $1.2\pm 0.4$ (-----) for the phase quadrature.}
\label{fig:optnoise}
\end{figure}

It is crucial to find how the interferometer noise, as expressed by the variance of the probe pulses, depends on the power of the probe laser. If the interferometer is shot noise limited, then according to Eq.~(\ref{eq:internoiseqe}) we would expect the noise to scale linearly with the probe power. Applying the same pulse sequence as described above, we first try to determine the noise in the amplitude quadrature. For that purpose, we block the probe arm of the interferometer and just observe the noise of the transmitted beam. In Fig.~\ref{fig:optnoise} a fit to the data shows that the amplitude noise depends linearly on the probe power. This shows that our detection is shot noise limited and in addition to that, we have found the shot noise level for later reference.

Next, we unblock the probe arm and look at the phase noise of the interferometer, which in the white light position should be dominated by shot noise. This is also shown in Fig.~\ref{fig:optnoise} and one sees that the dependence on the probe power is linear to within the uncertainty of the fit. On this, we conclude that the interferometer is shot noise limited in the probe power range 0.05 $\mathrm{\mu W}$ - 2 $\mathrm{\mu W}$. Moreover, Fig.~\ref{fig:optnoise} shows that the quantum noise is similar in the two quadratures, as one would expect from a coherent probe field.

We are now in a position to apply the interferometric detection to a sample of cold atoms. As a start we show how the interferometer can be used to nondestructively measure the number density of the trapped atomic cloud.


\subsection{Interferometry with cold atoms}\label{sec:dcphase}

\begin{figure}[t]
\begin{center}
\includegraphics[width=0.4\textwidth]{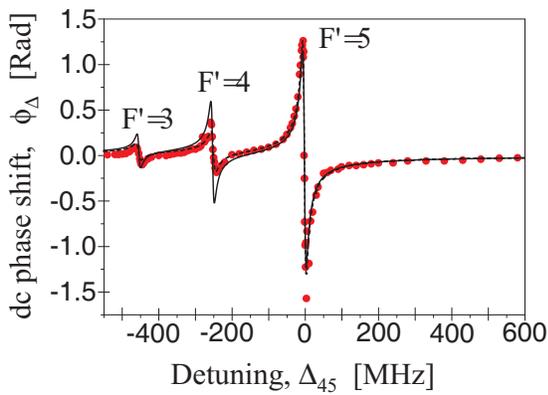}
\end{center}
\caption{(Color online) DC-phase shift due to dispersion by the atomic cloud with fit
(-----) to Eq. (\ref{eq:phaseshift}). A closer fit (-~-~-) is
obtained by letting the line strengths be independent, and thus not
given only by the value of the $6J$-symbols.}
\label{fig:dcphase}
\end{figure}

We begin with balancing the interferometer at half fringe in the absence of trapped atoms. Trapped atoms cause a DC-phase shift of the probe pulse. The measured signal must then be corrected for the absorption of the probe. From the corrected signal $i_{DC}$, the DC-phase shift $\phi_{\Delta}$ is deduced as $\phi_{\Delta}=\arccos{(i_{DC})}$. Measuring $\phi_{\Delta}$ as a function of the detuning, in the vicinity of $F=4$ level, we observe the dispersive behavior shown in Fig.~\ref{fig:dcphase}. Fitting the experimental data points using the Eq.~(\ref{eq:phaseshift}), we can determine the density of atoms
\begin{equation}\label{eq:numatoms}
 N_4 = \frac{2\pi C}{\lambda^2 l} =
 \frac{2\pi39.9}{(852~10^{-7})^2 ~0.1} = 4.3~10^{9}cm^{-3}\ ,
\end{equation}
where $C$ is the parameter given by the fit and is proportional to $\phi_0$. The estimated number of probed atoms is then
$5500$.

Notice that the theoretical fit on Fig.~\ref{fig:dcphase} overestimates the amplitude of the $6S_{1/2}(F\!=4) \rightarrow 6P_{3/2}(F'\!=3,4)$ transitions. This can be explained in the following way. We have accounted for the reduction in photon flux $\Phi$ and thereby the amplitude of the interference fringes as a result of absorption. However, we did not consider the optical pumping of the atoms due to the absorbed light, and the consequent reduction of the respective atomic density. The reduction of the atomic density at each resonance will depend mainly on the resonant term of the lineshape function $\mathcal{D}(\Delta)$, where the weight factor of the resonant terms should give the relationship between the absorption probabilities at the three resonances. However, for the $6S_{1/2}(F\!=4) \rightarrow 6P_{3/2}(F'\!=3,4)$ transitions we are pumping atoms over into the $F\!=3$ ground state, making them insensitive to the probe at the current detuning. On the other hand, the $6S_{1/2}(F\!=4) \rightarrow 6P_{3/2}(F'\!=5)$ transition is a cycling transition since the decay to the $F\!=3$ ground state is not allowed. Therefore, we do not depump the $F\!=4$ ground state at this resonance. Consequently, the fit to Eq.~(\ref{eq:phaseshift}) should avoid points close to the depumped $6S_{1/2}(F\!=4) \rightarrow 6P_{3/2}(F'\!=3,4)$ resonances, and the number $C$ used in Eq.~(\ref{eq:numatoms}) is indeed obtained from such a fit.

Closer theoretical calculations that also take stimulated emission and the branching ratios of the excited level decay into account, allow us to determine time evolution of the level populations for a given detuning and power of the probe. From this the average number of atoms that during the pulse have been pumped over into the $F\!=3$ ground state can be calculated, and used to estimate the reduction of the DC phase shift due to depumping. With this compensation applied to the data, the ratios between the amplitudes of the DC phase shift at the three resonances, come very close to values predicted by the transition strengths. 

We conclude that in the vicinity of the $F\!=4 \rightarrow F'\!=5$ transition the probe does not perturb the atomic population with our measurement conditions.


\subsection{Poisson noise of atomic population}\label{sec:spinnoise}

As we can see from Eq.~(\ref{eq:phaseshift}), at constant detuning the phase shift is proportional to the density of atoms $N$. Therefore, given a fixed detuning we can use the DC-phase shift as a measure for the number of atoms probed in the MOT~\cite{gibble,gibble1}. We vary the number of atoms trapped by varying the background Caesium pressure in the chamber.

The precision of our setup that we have so far reached does not enable us to measure at the detuning of $\Delta_{45}/2\pi=4312 \mathrm{MHz}$ where the atomic phaseshift cancels ($\phi_{\Delta}=0$) for equal populations in the two hyperfine ground states. With this in mind, the optimal choice of the probe detuning is dictated by the balance between the strength of the QND interaction and the strength of decoherence processes. As in Sec.~\ref{sec:spinsqueezing}, the signal refers to the photocurrent variance due to atoms and the noise is the photocurrent variance due to other noise sources. Eq.~(\ref{eq:signoise}) tells us that the signal to noise ratio $\kappa^2$ goes as the square of the detuning function $\mathcal{D}(\Delta)$ and linearly with the probe flux $\Phi$. From this relationship, we infer that a better signal is obtained at small detunings from the transitions where $\mathcal{D}(\Delta)$ is large. On the other hand, at small detunings the photon flux would decrease strongly due to absorption by the atoms. Moreover, the excitations can destroy the atomic coherences.

For the data presented here, the atomic population noise measurements were performed at the blue detuning of 15 MHz from the $F\!=4 \rightarrow F'\!=5$ transition. As a consequence we measure the noise of the atomic population in the $F=4$ ground state rather than the noise of the pseudospin component $j_z$. Using 2~$\mathrm{\mu s}$ long pulses of 0.6~$\mathrm{\mu W}$ power, this yields a pulse integrated rate of atomic transitions of $p_e = 15$. Obviously, this is far from being a QND measurement and the atomic coherence would have been completely lost under such measurement conditions. Normally, this would also mean that many atoms are transfered to the $F\!=3$ ground state, but since we are close to a cycling transition, virtually all of the atoms remain in the $F\!=4$ ground state after the measurement. Therefore, we have effectively relaxed the demand for the measurement to conserve the populations only, but not the coherences. However, the nature of the noise is unaffected by the measurement being or not being QND, and in that light the ability to measure the noise is a relevant indicator for the feasability of this procedure, in spite of the large amount of real transitions. 

Note that the photon flux is limited only by the detection, which is shot noise limited up to around $1~\mathrm{\mu W}$. With this photon flux of $5 \times 10^6$, every atom performs about 50 absorption cycles with the probe on resonance, whereby their recoil velocity reaches approximately 1.5~cm/s, which doesn't pose any problems on the timescales of our experiment.

For this measurement we employ a slightly more complex scheme of pulses. Each sequence consists of 3 pulses of a duration of 2 $\mathrm{\mu s}$ and equally separated by 10~ms. All three light pulses have the same 0.6~$\mathrm{\mu W}$ incident power that lies in the range where our detection is shot noise limited. Before each sequence we load the MOT and turn off the magnetic field just before the first of the three pulses. This means that the first pulse 1 probes the atomic cloud. With no trapping force to contain it, the MOT has long decayed at the arrival of the second pulse 2. In this way, the pulses 2 and 3 probe the chamber without any atomic cloud and are used as a reference. We repeat this sequence several thousand times to obtain the statistical information on the pulse areas $a_1$, $a_2$ and $a_3$. We subtract the pulse areas pairwise $d_{1-2} = a_1-a _2$ and $d_{2-3} = a_2 - a_3$, and calculate the mean values $\overline{d_{1-2}}$, $\overline{d_{2-3}}$ and variances $(\delta d_{1-2})^2$, $(\delta d_{2-3})^2$, over all the recorded sequences.

\begin{figure}[t]
\begin{center}
\includegraphics[width=0.4\textwidth]{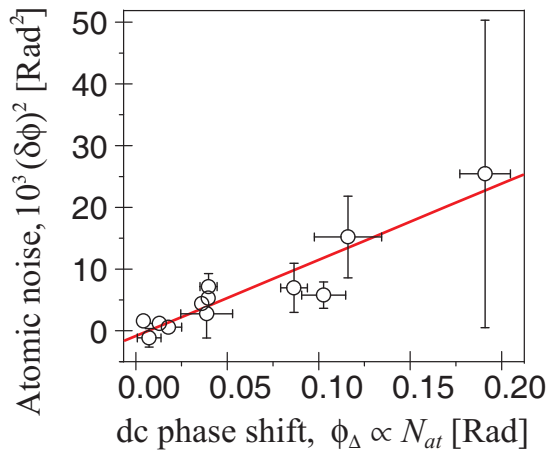}
\end{center}
\caption{(Color online) Phase noise induced in the probe light from the interaction with cold atoms. The
DC-phase shift is used as a measure for the number of atoms.}
\label{fig:spinnoise}
\end{figure}

Let us address the significance of these variances. For the first pulse the atoms are trapped in the MOT and thus, the pulse area variance $(\delta a_1)^2$ is given by Eq.~(\ref{eq:internoiseqe}) where, as mentioned earlier, we may neglect the classical phase noise $\mathcal{N}$ due to the white light alignment of the interferometer
\begin{equation}
(\delta a_1)^2 =  (\delta i)_p^2 +  \epsilon^2V\Phi^2 \left( \frac{\eta}{1+\eta}\right)^2(\delta \phi_{\Delta})^2 \cos^2(\phi_{\Delta})\ .
\label{eq:vara12}
\end{equation}
For atoms in a coherent superposition of two ground hyperfine state probed by light tuned in between the two states the atomic contribution to the phase noise $(\delta \phi_{\Delta})^2$ is induced by the quantum projection noise as in Eq.~(\ref{eq:atomnoise}). For atoms in a thermal state probed by light tuned close to one of the hyperfine transitions, as in our present setup, the atomic contribution to the phase noise is dominated by the population noise. The population noise arises from the statistical nature of the trapping process and from the motion of atoms in and out the probe region. As we demonstrate experimentally, it is characterized by Poission statistics with the width $\sqrt{N_{at}}$, same as for the projection noise.

For the last two pulses, recorded without any cold atoms ($\phi_{\Delta}=0$), there is no noise term from the atoms and consequently, the variance of the two areas may be written as
\begin{equation}
(\delta  a_j)^2 = (\delta i)_p^2 \quad,\quad j=2,3.
\label{eq:vara23}
\end{equation}
We see that one ought to be able to extract the shot noise and atomic noise from the appropriate pulse areas. What hinders it are slow thermal drifts of the interferometer that would dominate the variance if not corrected for. This is why we subtract the pulses pairwise to cancel the effect of the thermal drifts. Since the shot noise and atomic noise are uncorrelated it is easy to verify that for the subtracted areas $d_{1-2}$ and $d_{2-3}$ we have $(\delta d_{1-2})^2 = (\delta a_1)^2 + (\delta a_2)^2$ and $(\delta  d_{2-3})^2 = (\delta a_2)^2 + (\delta a_3)^2$ so that we can determine the shot noise and atomic noise contributions as
\begin{eqnarray}
(\delta i)_p^2 \!&=&\! \frac{(\delta d_{2-3})^2}{2} \nonumber ,\\[5pt]
(\delta \phi_{\Delta})^2 \!&=&\! \frac{(\delta d_{1-2})^2 -
(\delta d_{2-3})^2}{\epsilon^2V\Phi^2 \left( \frac{\eta}{1+\eta}\right)^2 \cos^2(\phi_{\Delta})}\ ,
\label{eq:dYpdphi}
\end{eqnarray}
respectively.

In Fig.~\ref{fig:spinnoise} we show the atomic contribution to the phase noise $(\delta \phi_{\Delta})^2$ as a function of the DC-phase shift $\phi_{\Delta}$, which, in turn, is proportional to the number of atoms. The linear fit, within the uncertainty, shows that the variance of the atomic fluctuations scale as $N_{at}$. Thus we conclude that the white light interferometry is capable of achieving the sensitivity at the level of projection noise fluctuations.


\section{Summary and Outlook}
\label{sec:outlook}

In this paper we propose a sequence of QND measurements and spin
rotations which allows to circumvent the projection noise limit of
accuracy of the Cs atom clock. The realization of this protocol
requires interferometric QND measurement of the atomic population
with the sensitivity at the projection noise level. Towards this
goal we demonstrate experimentally that  the shot noise limited
fiber optical interferometer at a white light setting can reach
the sensitivity sufficient to detect the projection noise under
conditions not far from those of the QND measurement.

The main parameter that can significantly improve the QND figure of merit,
 $\kappa^2$, is the resonant optical density $\alpha_0=\frac{\lambda^2lN}{2\pi}$, which is directly proportional to the number of atoms
 in the probing region. Increasing the optical density should allow to reduce the photon number and increase
 the detuning of the probe. Both those measures will improve the spin squeezed state preparation.
 The atomic density can be greatly increased compared to our present level by applying a far off resonant
 trap (FORT), where the dipole force from a focused laser beam traps the cold atoms~\cite{dipole}.
This naturally suggests that the QND scheme could be used in an
optical lattice clock where the number of atoms is preserved and
it is possible to have high densities. Given the density
limitation on the clock, an optimal density and atomic number
should be sought for, in order to improve the clock performance.

Another improvement lies in the optimization of the
interferometer, with the aim to reduce losses. This may mean
abandoning the fiber interferometer thereby practically
eliminating coupling losses ($\eta=1$). There is another reason
why the increased flexibility of a free space interferometer can
be an advantage. If the atoms are confined in a dipole trap, the
sample will be needle shaped, which will have a lensing effect on
the tightly focused probe beam. In this case it is desirable to be
able to include compensating lenses into the probe arm. The effect
of diffraction by atomic samples with different densities and
geometries on the spin squeezing is analyzed in detail in
~\cite{joerg}.

\begin{acknowledgments}
We are grateful to Dr. A. Lvovsky for providing us with the
schematics of the detectors used in this work. This research has
been supported by the Danish National Research Foundation and by
European networks CAUAC and QUICOV and partially funded
by the ESF-QIT program.
\end{acknowledgments}

\bibliographystyle{apsrev}
\bibliography{bibfile}

\begin{thebibliography}{31}
\bibitem[{foo({\natexlab{a}})}]{footnote2}
\bibinfo{note}{For detunings different from this, the optical phase will be
  more sensitive to one of the ground state populations, and therefore
  susceptible to noise in the atom number density.}
\end{thebibliography}
\end{document}